\documentclass[aps,prl,floatfix,twocolumn,footinbib]{revtex4-1}
\usepackage{graphicx,mathtools,amsmath,amsfonts,amssymb}

\begin{document}
    \title{Composite Particles and the Szilard Engine}
    \author{Tan Kok Chuan Bobby}
    \affiliation{Centre for Quantum Technologies, National University of Singapore, 3 Science Drive 2, 117543}
    \author{Dagomir Kaszlikowski}
    \affiliation{Centre for Quantum Technologies, National University of Singapore, 3 Science Drive 2, 117543}

    \begin{abstract}
        The Szilard engine is the simplest possible engine, composed only of one or more particles in a box. The box is then immersed in a heat bath and partitioned into two parts by a wall. It is known that in the cold temperature limit, one may extract more work out of elementary boson than out of elementary fermions. In this paper, we consider the amount of work that can be extracted out of a system of composite particles -- particles which are composed of two interacting elementary fermions of different species. We demonstrate that the amount of work extracted is closely tied to the amount of entanglement within the composite particles
    \end{abstract}\maketitle

    \textit{Introduction.} ---
 A Szilard engine \cite{Szilard1929} is an engine of fundamental importance in the field of thermodynamics. As a concept, it is useful to study the limitations of the laws of thermodynamics, and provides a bridge between the concepts of work extraction, and information about the system. The basic principles behind the Szilard engine is best explained by considering a particle trapped inside a box divided into 2 equal sections by a movable partitioning wall. This box is then placed inside a heat bath, and isothermal expansion is allowed to occur. In general, that particle can be on one side of the partition or the other, so if we completely have no knowledge of which side of the box the particle is in, then the average amount of work you can extract from the system is zero. However, suppose there is some hypothetical being (Maxwell's Demon)\cite{leff2003, maruyama2009} who is able to probe the location of the particle within the box. Such a being can then use his knowledge about the particle to extract nonzero amounts of work, in apparent violation of the second law of thermodynamics. Such a conclusion can be avoided if we assume that the demon has a limited amount of memory to store information about the particle, and that there is a minimal energy cost associated with the erasure of this information to make place for new ones \cite{brillouin1951,landauer1961,bennett1982} .

 One may break down the Szilard engine into a cyclic process consisting of 4 well defined procedures: (i)The insertion of the partition into the centre of the box. (ii) The measurement of the system. (iii) The moving of the partition through isothermal expansion. (iv) The removal of the wall which completes the cycle. The quantum version of this cycle had been previously studied for a general $N$ particle system, and the amount of work that is extractable from such a process is given by\cite{kim2011}:

 \begin{equation} \label{eqn:genwork}
    W_{Tot}=-k_BT\sum^{N}_{m=0}f_m\mathrm{ln} \left(\frac{f_m}{f^*_m}\right)
 \end{equation}

 Where $k_B$ is the Boltzmann constant,$T$ is the temperature of the heat bath, $f_m$ is the probability that during the measurement process we find $m$ particles to the left of the partition, and $f_m^*\coloneqq Z_m(l^m_{eq})/Z(l^m_{eq})$. $Z_m(l^m_{eq})$ is the partition function of the system when it has $m$ particles on the left, and the partition is at its equilibrium position $l^m_{eq}$, and  $Z(l^m_{eq})$ is the partition function of the system when the partition is at the position $l^m_{eq}$. The above assumes that the system is always in thermal equilibrium and that the particles may in general tunnel through the partition. In the special case where $N=2$, it is also known that:

 \begin{equation} \label{eqn:work}
    W_{tot}=-2k_BTf_0\mathrm{ln}(f_0)
 \end{equation}

 Where $f_0$ is the probability of finding both particles on one specific side of the partition during the measurement process. The case where $N=2$ is of particular significance because it is the simplest possible scenario where we can observe a divergence between the amount of extractable work due to the particle species. In the limit of $T\rightarrow 0$, $f_0=1/3$ for bosons,  $f_0=1/4$ for distinguishable particles, $f_0=0$ for fermions, clearly demonstrating different amounts of extractable work due to particle species. An intuitive explanation for this divergence is that in the low temperature limit, only the ground state of the system may be occupied. For fermions, the only acceptable ground state is one where there is one fermion on each side of the partition, due to Pauli's exclusion principle, while Bosons do not suffer the same restriction. Since no isothermal expansion can occur when one particle is on each side of the partition, you are able to extract more work from Bosons. It is worth noting that in the high temperature limit $T\rightarrow \infty $, $f_0\rightarrow \frac{1}{4}$, which is exactly the probability expected for classical distinguishable particles.

 In this paper, instead of considering elementary particles, we consider the amount of work extractable from a system of \emph{composite} particles. In particular, we consider composite particles composed of 2 tightly bound elementary particles, each belonging to a different species. The most common example of this is the hydrogen atom.

    \textit{Composite Particles.} --- Suppose we have two elementary fermions, described by their creation operators $a_n^{\dag}$ and $b_n^{\dag}$. These operators respect the anti-commutation relations $\{a_n,a_m^{\dag}\}=\delta_{n,m}$ and $\{b_n,b_m^{\dag}\}=\delta_{n,m}$ respectively, where $\{a,b\}\coloneqq{ab+ba}$. By interacting these two particles, we create a composite particle, described by the creation operator:

\begin{equation} \label{eqn:cdag}
    c^{\dag}\coloneqq \sum_n \sqrt{\lambda_n}a^{\dag}_n b^{\dag}_n
\end{equation}

Where the factors $\lambda_n$ are parameters that describe the internal structure and the amount of correlation within the composite particle. The form in Equation~\ref{eqn:cdag} is assured by the Schmidt decomposition of bipartite states\cite{law2005}. These creation operators do not in general behave like typical bosonic nor fermionic creation operators. For instance, the norm $\lVert (c^{\dag})^N|0\rangle \rVert^2$ is not equal to $N!$ as would be expected for bosonic operators, nor is it equal to zero for $n>1$, as would be expected for fermionic creation operators. As such, we define the factor $\chi_N$:

\begin{equation} \label{eqn:chi}
    \chi_N\coloneqq \lVert (c^{\dag})^N|0\rangle \rVert^2/N!
\end{equation}

One may interpret this $\chi_N$ factor in terms of the relative probability of success of creating the (unnormalized) state $(c^{\dag})^N|0\rangle$. To see this, consider the mixed state $\rho=\frac{1}{2}\left(|0\rangle\langle0|+\frac{1}{\chi_N N!}(c^{\dag})^N|0\rangle\langle0|(c)^N\right)$. For such a state, the system is equally likely to find itself in an $N$ particle state as a vacuum state. We now add a particle to the system, resulting in the state $\rho'=c^{\dag}\rho c / \mathrm{Tr}(c^{\dag}\rho c)$. One may then verify that following this operation, the probability of finding the system in the $N+1$ particle state $p_{N+1}$ relative to the probability the system is in the 1 particle state $p_1$ is simply

\begin{equation} \label{eqn:rprobn}
    p_{N+1}/p_1:=(N+1)\frac{\chi_{N+1}}{\chi_N}
\end{equation}

Which implies that you are more likely to add a composite particle to the $N$ particle state than an unoccupied state by a factor $(N+1)\frac{\chi_{N+1}}{\chi_N}$.

Suppose we are able to create composite particles occupying discrete energy states, which we label by the quantum number $p$ (we also refer to $p$ as the mode of the particle), so instead of a single creation operator, we have a family of creation operators $\{c^{\dag}_p\}$. Crucially, since we are dealing with identical composite particles, we assume that all the particle's internal structure is the same, so $c^{\dag}_p = \sum_n \sqrt{\lambda_n}a^{\dag}_{p,n} b^{\dag}_{p,n}$ for all possible values of $p$. Consider now 2 separate operations with $N$ particles: (i)where all $N$ particles are put into different unoccupied modes, and (ii) where all $N$ particles are placed in the same mode. By adding a particle one after another to the system, Equation~\ref{eqn:rprobn} suggests that the relative probability of achieving the operation (ii) relative to the operation (i) is given by:

\begin{equation} \label{eqn:rprob}
    (2\frac{\chi_2}{\chi_1})(3\frac{\chi_3}{\chi_2})\cdots (N\frac{\chi_N}{\chi_{N-1}})=N!\chi_N
\end{equation}

Where we see that the larger the value of $\chi_N$, the easier it is to successfully produce that state.

    \textit{A Semi-Classical Explanation for the Factor $\chi_N$}--- It is possible to provide an intuitive semi-classical explanation for the $\chi_N$ factor. Consider a set of creation operators ${c^{\dag}_p}$, such that $c^{\dag}_p\coloneqq \sum_n \sqrt{\lambda_n}a^{\dag}_{p,n} b^{\dag}_{p,n}$. We assume that the Hamiltonian of the system satisfies $\hat{H}(c_p^{\dag})^N|0\rangle=NE_p(c^{\dag})^N|0\rangle$, and that the number $n$ denotes the spatial coordinate of a particle in a one dimensional lattice. For such a system, an experimenter may, in principle, perform a measurement on the occupation number of a particular mode with energy $E_p$. This measurement is associated with a Hermitian operator which we denote $\hat{N}_p$. Note that the eigenstate of $\hat{H}$ is also an eigenstate of $\hat{N_p}$, so they commute.

We consider the eigenstate $(c_p^{\dag})^N|0\rangle$. Suppose the experimentalist measures first the occupation number, and subsequently measures the position of each of the $N$ composite particle (described by the quantum number $n$). His first measurement will tell him that he has $N$ particles with energy $E_p$, and his second measurement will allow him to infer the momentum of the particles, up to a sign factor, so he is able to determine the possible phase space coordinates of the system. We denote the possible measurement outcomes of the positional measurement to be $(n_1,n_2, ... ,n_N)$. For a system in contact with a heat bath, each possible state in phase space is attributed an equal \textit{a priori} probability if they have equal energy. However, for the quantum system under consideration, this assumption cannot be valid. For instance, consider the state of a single composite particle $c^{\dag}|0\rangle$. The probability of finding the particle in the position $n$ is $\lambda_n$, which in general is not equal for all $n$, which is incompatible with the assumption of equal \emph{a priori} probability for every possible state in phase space.

The way to resolve this is to associate each coordinate $n$ with some degeneracy $\Omega_n$ satisfying $\frac{\Omega_n}{ \sum_i \Omega_i}=\lambda_n$. One may think of this extra degeneracy as some classical hidden variable $\mu(n)$ describing the state system which may take $\Omega_n=\sum_{\mu(n)} \mu(n)$ possible different values for each coordinate $n$. This suggests that the complete state of the system may be described by $(\mu(n),n,E_p)$. If each of these states is attributed an equal \emph{a priori} probability, then a simple calculation shows that the probability of obtaining coordinate $n$ for a single composite particle is $\lambda_n$, as expected.

We may extend this to a system of $N$ composite particles in a relatively straightforward manner. For a quantum system the possible outcomes for the set of measurement coordinates $(n_1,n_2, ... ,n_N)$ may be assumed to ordered such that $n_1 < n_2 <...<n_N$, since:

 \begin{equation} \label{eqn:cdagn}
    (c_p^{\dag})^N|0\rangle = \sum_{n_1<n_2<...<n_N}N!\prod^N_{i=1}\lambda_{n_i}(a^{\dag}_{p,n_i}b^{\dag}_{p,n_i})|0\rangle
 \end{equation}

 This is due to the fact that $a^{\dag}_{p,n_i}$ and $b^{\dag}_{p,n_i}$ are identical fermions that obey Pauli's exclusion principle. Equation~\ref{eqn:cdagn} says that the probability of obtaining the measurement outcome $(n_1,n_2, ... ,n_N)$ is proportional to  the factor $\chi_N = \sum_{n_1<n_2<...<n_N}\lambda_{n_1}\lambda_{n_2}...\lambda_{n_N}$.

 Assume that $(n_1,n_2, ... ,n_N)$ satisfies $n_1 < n_2 <...<n_N$ as required by Pauli's exclusion, and that the degeneracy for each coordinate $n_i$ is $\Omega_{n_i}$. The result is that the total degeneracy for the state $(n_1,n_2, ... ,n_N)$ is given by $\Omega_{n_1} \times \Omega_{n_N} \times  ... \times \Omega_{n_N}$. Note the following relation:

\begin{equation}
    \begin{split}
        \chi_N &= \sum_{n_1<n_2<...<n_N}\lambda_{n_1}\lambda_{n_2}...\lambda_{n_N} \\
        &\propto \sum_{n_1<n_2<...<n_N}\Omega_{n_1}\Omega_{n_N} ... \Omega_{n_N}
    \end{split}
\end{equation}

Which is compatible with the assumption of equal \emph{a priori} probabilities. The factor $\chi_N$ may therefore be interpreted as a measure of the (hypothetical) degeneracy of a particular state.

    \textit{The Szilard Engine with 2 Composite Particles in the Low temperature limit.} --- We now consider a Szilard Engine composed of 2 composite particles, placed in some heat bath in the low temperature limit such that $T\rightarrow 0$. As previously mentioned, for elementary identical particles, the amount of work you can extract is intimately related to the effect of Pauli's exclusion principle. The situation for composite particles is more complex, as how closely they relate to a boson or a fermion depends on how the constituents of the particles interact with each other. Unlike elementary fermions, one cannot definitively rule out that 2 composite particles can occupy the same mode, since $\lVert (c^{\dag})^2|0\rangle \rVert^2 \neq 0$ so it is a valid state of the system. On the other hand, one cannot also immediately associate a composite particle with bosons because if the constituents of the composite particle are only very weakly correlated with each other, then $c^{\dag}$ is algebraically very similar to a fermionic creation operator, so observing two composite particle occupying the same mode must be a comparatively difficult and rare event.

We may account for this by considering the relative probability of the states under consideration. We first define the relative probability for a given set of occupation numbers, to be the following:

\begin{equation} \label{eqn:rprob2}
    P_{rel}(\{n_p\})\coloneqq (\frac{N!}{\prod_{p'} n_{p'}!})\prod_p n_p!\chi_{n_p}=N!\prod_p\chi_{n_p}
\end{equation}

Where $\{n_p\}$ is the set of occupation numbers $n_p$ for each mode $p$ and satisfies $\sum_p n_p =N$. Note that the term $(\frac{N!}{\prod_{p'} n_{p'}!})$ is the number of operations that can be performed by adding one particle at a time to produce the configuration $\{n_p\}$ of occupation numbers, while $\prod_p n_p!\chi_{n_p}$ is the relative probability of the operation as given by Equation~\ref{eqn:rprob}. Equation~\ref{eqn:rprob2} is therefore the product of the number of operations with the relative probability of success of said operations, as prescribed by Equation~\ref{eqn:rprob}.

For the Szilard engine with 2 particles, we only need to consider 3 possible states of the system: (a)One particle is on each side of the partition (b)both particles are to the left of the partition (c) both particles are to the right of the partition. We will assume that the energy of the system in all 3 cases are identical, and equal to some value $E_0$, corresponding to the ground state energy. In general, there is a possibility that the energy of configurations (b) and (c) may be slightly larger than that of (i). This is because the constituent particles are composed of fermions respecting Pauli's principle. By placing both of these composite particles in the same well, these fermions may experience exchange forces causing the mean separation between the particles to increase which, depending on the system, may in turn increase the observed energy of the state. We shall assume that this increase in energy due to exchange forces are negligible, and lies within the natural thermal fluctuations of the system. This assumption is valid in the regime where the dimensions of the trapping potential is macroscopic with respect to the microscopic composite particles. Intuitively, a macroscopic trapping potential should see only the composite particles, but not any effect on the energy of the system due to the microscopic internal structure of the particles.

Suppose the above physical conditions are satisfied, we can then apply Equation~\ref{eqn:rprob2} to (a),(b) and (c). For (a) There are two operations you can perform leading to 1 particle on each side to the partition: you may add one particle to the left side and followed by adding a particle to the right, or vice versa. Each of this operations correspond to a relative probability of 1, so according to Equation~\ref{eqn:rprob2}, its relative probability $P_{rel}(\{1,1\})$ is 2. For states (ii) and (iii) There is only one operation to perform, which is adding two particles one after another on one side of the partition, and this operation corresponds to the relative probability $2\chi_2$, so $P_{rel}(\{2,0\})=P_{rel}(\{0,2\})=2\chi_2$. This implies that the probability of finding 2 particles to the right of the partition is simply:

\begin{equation} \label{eqn:work2part}
    f_0=\frac{\chi_2}{1+2\chi_2}
\end{equation}

Thus the amount of extractable work (See Equation~\ref{eqn:work}) from the system increases with the factor $\chi_2$. Note that if $\chi_2=1$, as would be expected from an ideal boson, then $f_0=1/3$, and if $\chi_2=0$, as would be expected for ideal fermions, then $f_0=0$, so existing results regarding the elementary particles are retrieved. Interestingly, if $\chi_2=0.5$, the amount of work extractable from 2 classical distinguishable particles is obtained.  This allows us to employ $\chi_2$ as a measure of the bosonic or fermionic nature of composite particles. If $\chi_2>0.5$, then it is distinctively closer to being bosonic in nature, and if $\chi_2<0.5$, then it is closer to being a fermion. This observation that the factor $\chi_2$ is deeply related to the Bosonic and Fermionic properties of composite particles has previously been explored (See References~\cite{law2005, chudzicki2010, kurzynski2010, tichy2012}), and it now has an additional physical interpretation in terms of the amount of extractable work from a Szilard engine.

    \textit{Generalization to $N$ Composite Particles and General Temperature $T$}--- In this section, we generalize the above procedure for a Szilard Engine with $N$ composite particles and general temperature $T$. Note that the amount of extractable work as given in Equation~\ref{eqn:genwork}, is defined entirely by the partition functions of the system, generally given by $Z\coloneqq \sum_n e^{-\beta E_n}$ where $\beta$ is the inverse temperature $\frac{1}{k_B T}$, $E_n$ is the energy of the $n$th state of the system and $k_B$ is the Boltzmann constant. For the Szilard Engine, we may denote a particular state of the system by $(l,\{m_p\}_L,\{n_q\}_R)$, which is defined by the position of the partition $l$, and the set of occupation numbers $\{m_p\}_L$ and $\{n_q\}_R$ on the left and right of the partition respectively. The corresponding energy of the state is then $E(l,\{m_p\}_L,\{n_q\}_R)$. For elementary particles, one may then expect that the probability of finding the system in the state is proportional to $\exp\left(-\beta E(l,\{m_p\}_L,\{n_q\}_R)\right)$.

However, as per our previous conclusion, for composite particles, the probability of a given state of the system is also related to the set of occupation numbers of the state (See Equation~\ref{eqn:rprob2}). This implies that the probability of a given state $(l,\{n_p\}_L,\{m_q\}_R)$ with occupation numbers $\{m_p\}_L$ and $\{n_q\}_R$ is also proportional to the factor: $\prod_p \chi_{m_p} \prod_q \chi_{n_q}$. Thus, the expected probability of a state of the system should $p(l,\{m_p\}_L,\{n_q\}_R)$ satisfies the following:

 \begin{equation}
    \begin{split}
        p&(l,\{n_p\}_L,\{m_q\}_R) \\
        &\propto \prod_p \chi_{n_p} \prod_q \chi_{m_q} \exp\left(-\beta E(l,\{n_p\}_L,\{m_q\}_R)\right) \\
        &\coloneqq Z(l,\{m_q\}_L,\{n_p\}_R)
    \end{split}
 \end{equation}

 This allows us to redefine the necessary partition functions to calculate the extractable work from composite particles. For a given position of the partition $l$, the partition function of the system with $m$ composite particles to the left of the partition , $Z_m(l)$, is given by:

 \begin{equation} \label{eqn:part1}
        Z_m(l)= \sum_{(\{m_q\}_L,\{n_p\}_R) \in M} Z(l,\{m_q\}_L,\{n_p\}_R) \\
 \end{equation}

where the set $M$ is defined to be the set of occupation numbers $(\{m_q\}_L,\{n_p\}_R)$ satisfying $\sum_q m_q=m$ and $\sum_p m_p =N-m$. The total partition function $Z(l)$ is then further given by:

\begin{equation} \label{eqn:part2}
    Z(l)=\sum_{m=0}^{N}Z_m(l)
\end{equation}

Where the partition function $Z(l)$ in general depends on the system being studied. With the expressions given by Equations \ref{eqn:part1} and \ref{eqn:part2}, it is then in principle possible to compute $f_m$ and $f_m^*$ which, together with the temperature of the bath, fully defines the amount of extractable work for a Szilard Engine composed of $N$ composite particles at temperature $T$, as given by Equation~\ref{eqn:genwork}. Note that $1-N(1-\chi_2)\leq \frac{\chi_{N+1}}{N} \leq \chi_2$~\cite{chudzicki2010}, and as a result the partition functions for bosons and fermions are retrieved in the limits $\chi_2\rightarrow 1$ and $\chi_2\rightarrow 0$ respectively.

    \textit{Example: The Hydrogen Atom} --- We illustrate our results by considering a Szilard Engine composed of 2 Hydrogen Atoms in the low temperature limit. The Hydrogen atom has already been extensively studied and its detailed properties are well known. In particular, it is known that the purity of a Hydrogen atom confined within a harmonic oscillator potential is given by\cite{chudzicki2010}:

\begin{equation} \label{eqn:hpurity}
    P=\frac{33}{4\sqrt{2\pi}}(\frac{a_0}{b})^3
\end{equation}

Where $a_0$ is the Bohr radius, $b^2\coloneqq\frac{\hbar}{m \omega}$, $\hbar$ is the Planck constant, $m$ is the mass of the hydrogen atom, and $\omega$ is the natural frequency of the harmonic oscillator. In general, the parameter $b$ increases with with the size of the trap, so the purity decreases as the dimensions of the trap increases. One may then compute the amount of extractable work in the low temperature limit by applying Equation~\ref{eqn:work2part}, noting that $\chi_2=1-P$. For a trapping frequency of the order of 10kHz, one can verify that the purity is approximately of the order of $10^{-13}$, so the amount of work extractable from hydrogen atoms is effectively the same as for ideal bosons for small systems.

    \textit{Conclusion} --- In summary, we have studied the amount of work that is extractable from composite particles using the the quantum analogue of a Szilard Engine. The behaviour of composite particles is deeply related to the factors $\chi_N$ that appear as a result of the internal structure of the composite particles. It is worth noting that $\chi_2$ in particular may be considered as a measure of the entanglement present within the composite particle~\cite{law2005}. These normalization factors have previously been studied within the context of bosonic behaviour of composite particles, and it is known that as $\chi_2 \rightarrow 1$, or in the limit of infinite entanglement within the composite particle, the creation operators of the composite particles begin to adopt many of the algebraic properties of bosonic operator \cite{law2005,chudzicki2010,ramanathan2011,combescot2001}. Here, we show that the same factors arise naturally from physical considerations in the context of work extraction. Through semi-classical reasoning, we also demonstrate that these factors may be considered as a measure of the amount of degeneracy for a given state.

Of particular interest is a Szilard engine with 2 composite particles in the low temperature limit. It is under these conditions that the difference between fermions and bosons, as characterised by the amount of work you can extract from them, is the most stark. We demonstrate that composite particles may bridge the gap between fermions and bosons, forming a continuum of particle species characterized by $\chi_2$ that approaches fermionic behaviour as $\chi_2 \rightarrow 0$, and approaches bosonic behaviour as $\chi_2 \rightarrow \infty$. Finally, we also provide generalizations to the $N$ particle Szilard engine at general temperature $T$

This research is supported by the National Research Foundation and Ministry of Education of Singapore.

    \bibliographystyle{apsrev}
    \bibliography{references}

\end{document}